\documentclass[aps,prb,twocolumn,showpacs,superscriptaddress,floatfix]{revtex4-1}

\usepackage{amsfonts,amsmath,graphicx}

\newcommand{\para}{{\slash{}\!\slash{}}}
\newcommand{\Exp}[1]{\mathrm{e}^{#1}}

\begin{document}

\title{Macroscopic quantum tunneling in quartic and sextic potentials:\\application to a phase qubit}
\author{N.~Didier}
\affiliation{Scuola Normale Superiore, NEST and Istituto Nanoscienze-CNR, Pisa, Italy}
\author{F.~W.~J.~Hekking}
\affiliation{Laboratoire de Physique et de Mod\'elisation des Milieux Condens\'es, Universit\'e Joseph Fourier and CNRS, BP 166, 38042 Grenoble, France}

\begin{abstract}
Macroscopic quantum tunneling of the phase is a fundamental phenomenon in the quantum
dynamics of superconducting nanocircuits. The tunneling rate can be controlled in such
circuits, where the potential landscape for the phase can be tuned with different external
bias parameters. Precise theoretical knowledge of the macroscopic quantum tunneling rate
is required in order to simulate and understand the experiments. We present a derivation,
based on the instanton technique, of an analytical expression of the escape rate in
general quartic and symmetric sextic potentials comprising two escape paths. These new
potentials were recently realized when creating a noise-insensitive phase qubit in the
camel-back potential of a dc SQUID.
\end{abstract}
\maketitle

\section{Introduction}

Due to the enormous progress in nanotechnology, superconducting nanocircuits are
nowadays commonly manipulated in the quantum regime~\cite{reviewBuisson,reviewGuichard}. The collective behavior
of superconducting electrons on the macroscopic scale is described with the
superconducting phase of the Cooper pair condensate. As a function of this collective
degree of freedom, the relevant potential energy for the phase is composed of minima
separated by potential barriers. At sufficiently low temperatures, the quantization of
energy in the potential wells generates quantum states that can be used in quantum
information processing. The quantum dynamics as well as the readout of the quantum state
are then governed by the tunneling between different minima of the potential. It is thus
crucial to know the appropriate macroscopic quantum tunneling (MQT) rate of the phase for
an arbitrary potential shape. Although this quantity can be calculated numerically in a
general one- or multi-dimensional potential, it is much more convenient to have an
analytical solution when possible. In the following, we show that for a general quartic
potential and a symmetric sextic potential it is still possible to derive an analytical
solution for the MQT rate. We will take a phase qubit as a physical example~\cite{reviewMartinis},
but our results apply to any system presenting such potentials.

Usually, the precise shape of the potential depends on certain bias parameters. Most of
the experiments on superconducting circuits are performed with current-biased Josephson
junctions. The relevant potential for the phase is then the so-called tilted washboard
potential~\cite{Likharev}. If the bias current is close to the critical current, the
washboard potential is well approximated by a cubic polynomial with a local potential
minimum from which the phase can escape into a continuum by tunneling. However,
interesting features appear away from this working point. In case of a SQUID loop for
instance, at small current bias and with a flux bias close to half a flux quantum, the
quantum states become weakly sensitive to current noise. In this region, the relevant
potential has a quartic contribution and presents two escape points. This working point
was used in the experiment of Ref.~\onlinecite{camelback}, where a phase qubit was
created in the so-called camel-back potential of a dc SQUID.
For this experiment, the results of the present paper can be used for simulations and data processing~\cite{these}.

In this paper we derive the expression of the MQT rate in a general quartic potential
from which tunneling occurs through two possible escape paths, as depicted in
Fig.~\ref{figquartic}. We use the instanton formalism developed in
Ref.~\onlinecite{Coleman}. We also give the MQT rate for a symmetric sextic potential.
The analytical expression is Eq.~\eqref{GLRgenpot} and the total escape rate turns out to
be the sum of the escape rates for tunneling through each barrier. We find a simple
result in the case of the experimentally relevant camel-back potential, Eq.~\eqref{Gcb},
depending on the bottom well frequency and the barrier height. We finally discuss the
detailed shape of the potential in terms of the experimental parameters of the phase
qubit in the two limiting cases of the washboard and the camel-back potential in
Appendix~\ref{apptopo}.

\section{Quartic potential in a SQUID}
\label{phasequbit}

\begin{figure}[t]
   \includegraphics[height=6.5cm]{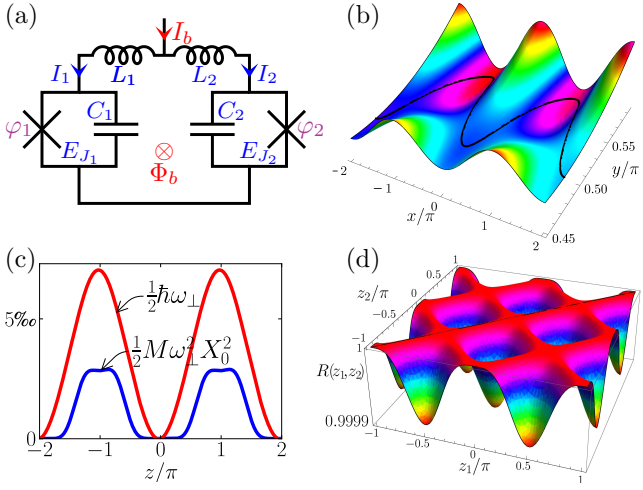}
   \caption{(a) Schematic representation of the SQUID.
   (b) Potential of the SQUID $u(x,y)$ and the path of minimum curvature connecting the minima through the saddle points.
   (c) Additional potentials $\frac{1}{2}\hbar\omega_\bot$ and $\frac{1}{2}M\omega_\bot^2X_0^2$ in units of the barrier height along the path of minimum curvature, the constant contribution has been removed.
   (d) Overlap of the ground state wave functions in the transverse direction as a function of the two abscissas $z_1$ and $z_2$.}
   \label{figsquid}
   \end{figure}

The SQUID is a superconducting loop interrupted by two Josephson
junctions~\cite{tinkham}, see Fig.~\ref{figsquid}.a. The Josephson junctions are
characterized by their critical currents $I_{c_{1,2}}$ or equivalently their Josephson
energies $E_{J_{1,2}}=\frac{\hbar}{2e}I_{c_{1,2}}$, their capacitances $C_{1,2}$, and the
phase differences $\varphi_{1,2}$ across them. Each arm of the loop has a self-inductance
$L_{1,2}$ and carries a partial current $I_{1,2}$. The SQUID is biased by the current
bias $I_b=I_1+I_2$ and the flux bias $\Phi_b$. We introduce the reduced variables
$x=\tfrac{1}{2}(\varphi_1+\varphi_2)$ and $y=\tfrac{1}{2}(\varphi_1-\varphi_2)$. In the
following we consider a SQUID for which the capacitances in each arm are almost equal,
$C_1\simeq C_2$. The Hamiltonian in the $(x,y)$ plane reads
$H=\tfrac{1}{2C}(Q_x^2+Q_y^2)+E_Ju(x,y)$, where $Q_x$ and $Q_y$ are the charges conjugate
to the phases $x$ and $y$, $C^{-1}=C_1^{-1}+C_2^{-1}$ is the total inverse capacitance,
$E_J=E_{J_1}+E_{J_2}$ is the total Josephson energy. The characteristic frequency of the
phase dynamics is the plasma frequency $\omega_p=\sqrt{8E_JE_C}/\hbar$, where
$E_C=e^2/2C$ is the charging energy of the SQUID. Due to the symmetry in the capacitances
there is no coupling between the charges $Q_x$ and $Q_y$. We define the reduced biases
$s=I_b/I_c$, $y_b=\Phi_b/2\phi_0$ with the reduced flux quantum $\phi_0=\hbar/2e$, the
critical current asymmetry $\alpha=(I_{c_2}-I_{c_1})/I_c$, the loop inductance asymmetry
$\eta=(L_2-L_1)/L$, and the junction-to-loop inductance ratio $b=2\phi_0/LI_c$, where
$I_c=I_{c_1}+I_{c_2}$ is the total critical current and $L=L_1+L_2$ is the total
inductance. The reduced potential then reads
   \begin{equation}
   u(x,y)=-\cos x\cos y-\alpha\sin x\sin y-sx-s\eta y+b(y-y_B)^2.
   \label{potentialxy}
   \end{equation}
The dynamics is then equivalent to a fictitious particle of mass $M=\phi_0^2C$ and
coordinates $(x,y)$ moving in the potential $u(x,y)$.

The typical shape of the potential $u(x,y)$, Eq.~\eqref{potentialxy}, is composed of
local minima and saddle points connected by narrow valleys (see Fig.~\ref{figsquid}.b).
The presence of extremal points is governed by the bias fields. Indeed, if $s$ is larger
than unity for example, there are no minima. For a given value of the flux, we define the
critical current as the value $s_c$ at which the minima merge with the saddle points,
consequently there are neither minima nor saddle points for bias currents $s$ larger than
$s_c$. The diagram $s_c(y_B)$ is called the critical diagram.

To study the phase dynamics from minimum to minimum in this potential, we will suppose
that the fictitious particle follows the path of minimal curvature. We can then describe
the dynamics with one variable, since the path in the plane $(x,y)$ is determined by $x$
and $y=\zeta(x)$ (see Fig.~\ref{figsquid}.b). We call $z(x)$ the curvilinear abscissa
along the path of minimum curvature, defined by
   \begin{equation}
   z(x)=z_0+\int_{x_0}^x\mathrm{d}x'\sqrt{1+\zeta'^2(x')}.
   \end{equation}
It is now tempting to consider the motion according to the effective one-dimensional potential $E_Ju(z)$.
In the following we check that the perpendicular dynamics does not generate additional terms.

Let us consider a point $(x,y)$ at abscissa $z$ on this path and call $\theta$ the angle
between the $x$ axis and the direction of minimal curvature ($\theta\in[-\pi,\pi]$)
   \begin{equation}
   \theta(x)=\frac{1}{2}\arctan\!\left(\frac{2\partial_{xy}^2u(x,y)}{\partial_{xx}^2u(x,y)-\partial_{yy}^2u(x,y)}\right).
   \end{equation}
The direction of minimal curvature is denoted by $X_\para$ and the perpendicular
direction, given by the angle $\theta+\pi/2$, is denoted by $X_\bot$. We can then express
the Hamiltonian in terms of $X_\para$, $X_\bot$, $P_\para=M\dot{X}_\para$, and
$P_\bot=M\dot{X}_\bot$ by performing a rotation of angle $\theta$.

In the transverse direction, the frequency
$\omega_\bot=\sqrt{\partial_{X_\bot}^2u(x,y)}\,\omega_p$ is much larger than the bottom
well frequency $\omega_0$ in the direction of minimum curvature
($\omega_\bot/\omega_0\simeq7$ in the experiment of Ref.~\onlinecite{camelback}). The
perpendicular dynamics is thus much faster than the dynamics along the path of minimum
curvature, and can be averaged out. The dynamics in the longitudinal direction is then
governed by the effective Hamiltonian $H_\para=\langle H\rangle_\bot$ obtained after
averaging $H$ over the transverse degree of freedom, the kinetic part being obtained
using $P_x^2+P_y^2=P_\para^2+P_\bot^2$. To proceed, we consider an adiabatic evolution in
the longitudinal direction and use $X_\para$ as a fixed parameter. We suppose the
perpendicular dynamics frozen in the ground state of an effective potential of bottom
well frequency $\omega_\bot$. The effective perpendicular dynamics, obtained within the
harmonic approximation, is governed by the Hamiltonian
   \begin{equation}
   H_\bot=\tfrac{1}{2M}P_\bot^2+\tfrac{1}{2}M\omega_\bot^2(X_\bot-X_0)^2-\tfrac{1}{2}M\omega_\bot^2X_0^2,
   \label{habot}
   \end{equation}
where the parameter $X_0$ originates from the slope of the potential in the transverse
direction. Averaging over the motion in the transverse direction gives rise to two
contributions to the potential of $H_\para$: $\langle
H_\bot\rangle_\bot=\tfrac{1}{2}\hbar\omega_\bot-\tfrac{1}{2}M\omega_\bot^2X_0^2$. In
practice, however, this additional potential $\langle H_\bot\rangle_\bot/E_J$ is only a
small correction to $u(z)$ (see Fig.~\ref{figsquid}.c) and will be neglected in the
following. The quantum dynamics of the phase can thus be treated in the one-dimensional
potential $\mathcal{U}(z)=E_Ju(z)$ following the path of minimum curvature.

To quantify the adiabaticity, we evaluate the overlap of the transverse ground-state wave functions at two points $z_1$ and $z_2$ along the path of minimal curvature.
It turns out to be equal to $R(z_1,z_2)=\frac{\sqrt{2}\left[\omega_\bot(z_1)\omega_\bot(z_2)\right]^{1/4}}{\left[\omega_\bot(z_1)+\omega_\bot(z_2)\right]^{1/2}}$.
If we choose $z_2=z+\delta z$ close to $z_1\equiv z$, we have
$R(z,z+\delta z)\simeq1-\left(\frac{\omega'_\bot(z)\,\delta z}{4\omega_\bot(z)}\right)^2$.
In order to obtain an adiabatic evolution in the ground state, the condition is thus
$\displaystyle \frac{\omega_\bot(z)}{\omega'_\bot(z)}\gg\delta z,$
where the characteristic distance $\delta z$ is of the order of unity ($\delta z<2\pi$).
The numerical evaluation of these quantities shows that the adiabaticity condition is satisfied (see Fig.~\ref{figsquid}.d).

   \begin{figure}[t]
   \centering
   \includegraphics[height=6.5cm]{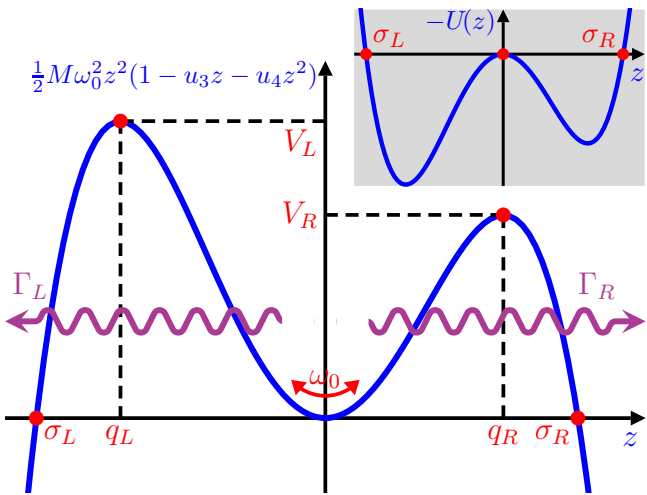}
   \caption{General quartic potential $U(z)=\frac{1}{2}M\omega_0^2z^2(1-u_3z-u_4z^2)$, comprising two escape paths at $\sigma_R$ and $\sigma_L$, passing through the barriers with a height $V_{L,R}$ at the abscissa $q_{L,R}$.
   The inset is the inverted potential obtained after a Wick rotation.}
   \label{figquartic}
   \end{figure}

We choose the origin of the curvilinear abscissa along the path of minimum curvature at a particular minimum $\mathrm{Min}_C$ with coordinates $(x_m,y_m)$.
Along the path of minimum curvature $(x,\zeta(x))$, this minimum is connected to two saddle points $\mathrm{Saddle}_{L,R}$, which are both connected to another minimum $\mathrm{Min}_{L,R}$.
The shape of the barriers is given by the one-dimensional potential $\mathcal{U}(z)$.
The convention is to call $\mathrm{Min}_R$ the peripheral minimum with the lowest potential energy and to orient $z$ towards $\mathrm{Min}_R$.
If we focus on the dynamics around the minimum, this potential is well approximated by its Taylor expansion up to fourth order, \textit{i.e.} $\mathcal{U}(z) = \,\mathcal{U}(0) + U(z) + O(z^5)$ with
   \begin{equation}
   U(z)=\tfrac{1}{2}M\omega_0^2z^2\left(1-u_3z-u_4z^2\right),
   \label{genquarticpot}
   \end{equation}
where the bottom well frequency $\omega_0\sim\omega_p$ and the coefficients $u_3\geq0$, $u_4\geq0$ are obtained numerically or with approximate expressions in specific cases (see Appendix~\ref{apptopo}).
The abscissa $q_{L,R}=z(\mathrm{Saddle}_{L,R})$ of the saddle points are
   \begin{equation}
   q_{L,R}=\Big(-3u_3\mp\sqrt{9u_3^2+32u_4}\Big)/8u_4.
   \end{equation}
We also define the escape points at abscissa $\sigma_{L,R}$ where $U(\sigma_{L,R})=0$
   \begin{equation}
   \sigma_{L,R}=\Big(-u_3\mp\sqrt{u_3^2+4u_4}\Big)/2u_4.
   \end{equation}
The point at $\sigma_R$ is always defined whereas $\sigma_L$ has a physical meaning only when $U(\mathrm{Min}_L)\leq U(\mathrm{Min}_C)$.
When the coefficient $u_4$ is larger than $u_3$, the potential is composed of two barriers.
Because of these two ``humps'' (see Fig.~\ref{figquartic}), this potential has been called the camel-back potential~\cite{camelback}.

The dynamics in a given minimum is governed by the Hamiltonian
   \begin{equation}
   H_\mathrm{aho}=\hbar\omega_0\left(P_z^2+Z^2\right)-a\hbar\omega_0Z^3-b\hbar\omega_0Z^4,
   \label{Haho}
   \end{equation}
where $Z=\sqrt{M\omega_0/\hbar}\,z$ is the reduced position operator, $P_z$ is the corresponding momentum operator, $a=u_3\sqrt{\hbar/M\omega_0}/2$, and $b=u_4\hbar/2M\omega_0$.
Due to the presence of the terms proportional to $a$ and $b$, Hamiltonian~\eqref{Haho} describes an anharmonic oscillator.
At sufficiently low temperatures below the plasma frequency, the energy spectrum of the particle is quantized in the minimum.
Assuming the anharmonicity to be weak, a second order perturbation theory in $a$ and $b$ leads to a transition frequency between the levels $n$ and $n-1$
   \begin{equation}
   h\nu_{n,n-1}=\hbar\omega_0\left(1-n\Lambda_n\right),
   \end{equation}
where $\Lambda_n$ is the anharmonicity of the oscillator
   \begin{equation}
   \Lambda_n=3b+\tfrac{15}{4}a^2+\tfrac{3}{4}b^2(17n+7/n).
   \end{equation}
The anharmonicity depends on the working point $(I_b,\Phi_b)$.
A sufficiently large anharmonicity is necessary to reach the two-level limit.
Indeed, the manipulation of the states is performed with a microwave in resonance with the energy difference between the ground state and the first excited state and
for a harmonic oscillator, where the anharmonicity vanishes, this microwave would excite all the levels.
For a sufficiently large $\Lambda$, the two states $|0\rangle$ and $|1\rangle$ are decoupled from the others and constitute a qubit.
The confining potential depending on the phases, this two level system is called a phase qubit~\cite{claudonbalestro,coopermartinis,lisenfeld,dutta}.

To understand the quantum dynamics of the phase qubit in the camel-back potential, comprising two barriers, it is necessary to calculate the tunneling rate in a general quartic potential Eq.~\eqref{genquarticpot}.
We present the derivation of the escape rate in the following section using the instanton technique.
We also extend the result in the case of a sextic contribution when the potential is symmetric.

\section{Derivation of the macroscopic quantum tunneling rate}

The expression of the escape rate in the quartic potential sketched in Fig.~\ref{figquartic}, with a double escape path, is not explicitly given in the literature of MQT (see \textit{e.g.} Refs.~\onlinecite{Kagan,razavy,takagi}) and has to be calculated.
We use the instanton technique~\cite{Callan,Coleman} to derive the escape rate from the metastable state of the quartic potential Eq.~\eqref{genquarticpot}.

The escape rate $\Gamma$ is obtained from the quantum ground state of the fictitious particle confined in the well.
Indeed, if $E_0$ is the ground state energy, the norm of the particle wave function $\Psi(z,t)$ in the well follows the temporal evolution
   \begin{equation}
   \int\mathrm{d}z\left|\Psi(z,t)\right|^2=\Exp{2 \textrm{Im}\,E_0 t/\hbar},
   \end{equation}
which decays exponentially with the decay rate
   \begin{equation}
   \Gamma=-\frac{2}{\hbar}\,\textrm{Im}\,E_0.
   \end{equation}
The existence of a non-zero imaginary part comes from the possibility to tunnel out of the trap.
The lifetime of the state is equal to $\Gamma^{-1}$.
The quantum dynamics of the particle is governed by the Hamiltonian $\displaystyle H=\tfrac{1}{2M}P^2+U(z)$ where the potential $U(z)$ comprises a local but not global minimum, fixed at $z=0$ for convenience.
If we were dealing with a classical particle, the equilibrium state would be the particle at rest at the local minimum.
This is however not possible in the presence of quantum fluctuations.
Moreover, quantum mechanically, the extension of the wave function allows the particle to evolve out of the trap where the potential energy is lower.
The classical equilibrium is thus a ``false ground state''~\cite{Callan}.
The energy of the particle in the well is calculated from the propagator
   \begin{equation}
   G(t)=\langle0|\Exp{-iHt/\hbar}|0\rangle=\Exp{-iE_0t/\hbar},
   \label{defpropagator}
   \end{equation}
where $|0\rangle$ is the position eigenstate at $z=0$.

The probability amplitude Eq.~\eqref{defpropagator} is obtained after integrating over all possible trajectories $z(t)$ satisfying the boundary conditions $z(t_i)=z(t_f)=0$.
For a given path, the phase of the contribution is the corresponding action in units of the action quantum $\hbar$~\cite{FeynmanHibbs}.
To proceed, let us perform first a Wick rotation: time $t$ is replace by the imaginary time $\tau=it$.
Then the action for a fixed trajectory becomes the Euclidean action
   \begin{equation}
   S(z)=\int_{-T/2}^{T/2}\textrm{d}\tau\left[\frac{M}{2}\dot{z}^2+U(z)\right],
   \end{equation}
where the dot means time derivative with respect to $\tau$ and $it_{i,f}=\mp T/2$.
The effect of the Wick rotation on the action is thus to invert the potential (see Fig.~\ref{figquartic}, inset).
In other words, in the part of space between the minimum and the exit point, the momentum of the particle is imaginary: $P^2=-2MU(z)<0$.
But if we choose the time variable $t$ as an imaginary variable $\tau=it$, then the motion is possible in the sense of classical dynamics in the inverted potential $-U(z)$.
In terms of path integrals the propagator reads
   \begin{equation}
   G(T)=\mathcal{N}\int_{z(-T/2)=0}^{z(+T/2)=0}\mathcal{D}z\,\Exp{-S(z)/\hbar}.
   \label{euclideanactionpathintegral}
   \end{equation}
The parameter $\mathcal{N}$ is a normalization factor and $\mathcal{D}z$ denotes the integration over all functions $z(t)$ obeying the boundary conditions.
The propagator~(\ref{euclideanactionpathintegral}) can be evaluated in the semiclassical limit, where the functional integral is dominated by the stationary point $\bar{z}$ of $S(z)$.
The  stationary point satisfies $\displaystyle\left.\delta S(z)/\delta z\right|_{\bar{z}}=0$, \textit{i.e.},
   \begin{equation}
   -M\ddot{\bar{z}}+U'(\bar{z}) = 0.
   \label{eqforbounce}
   \end{equation}
The action of the stationary point is $\displaystyle S_0=\int_{-T/2}^{T/2}\textrm{d}\tau\left[\tfrac{M}{2}\dot{\bar{z}}^2+U(\bar{z})\right]$.
The path integral can be evaluated with the method of steepest descent around the stationary trajectory, where the fluctuations $\tilde{z}$ can be expressed on the eigenfunctions $z_n$ of the operator $-M\partial_t^2+U''(\bar{z})$, $\tilde{z}=\sum_n\xi_nz_n$.
The functional integral, with $\mathcal{D}\tilde{z}=\prod_n(2\pi\hbar)^{-\frac{1}{2}}\mathrm{d}\xi_n$, gives rise to a functional determinant,
   \begin{multline}
   \int_{z(-T/2)=0}^{z(+T/2)=0}\mathcal{D}z\,\Exp{-S(z)/\hbar}=\Exp{-\omega_0T/2-S_0/\hbar}\\\times\left(\mathrm{det}\!\left[-M\partial_t^2+U''(\bar{z})\right]\right)^{-\frac{1}{2}}.
   \end{multline}
The prefactor $\mathcal{N}$ is determined using the Gelfand-Yaglom formula, establishing that
the ratio of two determinants of the same particle with the same energy in two different potentials is equal to the ratio of the corresponding wave functions.
Using the asymptotic expression of the wave function of a harmonic oscillator of mass~$M$ and frequency~$\omega_0$ in the ground state, $\displaystyle \psi_0(T/2)\sim\Exp{\omega_0T}/2\omega_0$, we get
   \begin{align}
   G(T) &= \sqrt{\frac{M\omega_0}{\pi\hbar}}\,\Exp{-\omega_0T/2-S_0/\hbar} K(T),\\
   K(T) &= \left(\frac{\mathrm{det}\!\left[-M\partial_t^2+U''(\bar{z})\right]}{\mathrm{det}\!\left[-M\partial_t^2+M\omega_0^2\right]}\right)^{-\frac{1}{2}}.
   \end{align}

Let us specify the shape of the trajectory $\bar{z}(\tau)$, solution of
Eq.~\eqref{eqforbounce}, to determine the action $S_0$ and the function $K(T)$. We
consider the nontrivial solutions where the particle can start at the top of the hill in
the inverted potential, bounces off the potential on the right at $z=\sigma_R$ or on the
left at $z=\sigma_L$ and returns to the top of the hill. For $T\to\infty$, we call this
trajectory the right, and respectively the left, ``bounce''~\cite{Coleman}. The bounce
has an energy $E_0=0$, thus
   \begin{equation}
   \dot{\bar{z}}=\sqrt{\frac{2}{M}U(\bar{z})},
   \label{eqbounceorbit}
   \end{equation}
and the bounce action reads
   \begin{equation}
   S_{0,j}=2\int_{0}^{\sigma_j}\textrm{d}z \sqrt{2MU(z)},
   \end{equation}
where the index $j$ stands for ${L,R}$.
To evaluate the functional determinant in $K_j(T)$, we follow the general framework of Ref.~\onlinecite{Coleman}.
The lowest eigenvalues have to be treated separately.
First, noticing that $M\dddot{\bar{z}}=U''(\bar{z})\dot{\bar{z}}$, we find that $z_2=\sqrt{M/S_{0,j}}\dot{\bar{z}}$ is an eigenvector of the operator $-M\partial_t^2+U''(\bar{z})$ with a vanishing eigenvalue $\lambda_2$.
This eigenvalue is excluded from the determinant, now noted with a prime, and the integration over $\xi_2$ yields a prefactor $\sqrt{S_{0,j}/2\pi\hbar M}T$.
Second, as $z_2$ has a node, $\lambda_2$ is not the lowest eigenvalue.
There is a negative eigenvalue, $\lambda_1$, and the Gaussian integral over $\xi_1$ leads to $i/2\sqrt{|\lambda_1|}$.
The intermediate expression for $K_j(T)$ is then
   \begin{equation}
   K_j(T)=\frac{i}{2}\sqrt{\frac{S_{0,j}}{2\pi\hbar M}}\left|\frac{\mathrm{det}'\!\left[-M\partial_\tau^2+U''(\bar{z})\right]}{\mathrm{det}\!\left[-M\partial_\tau^2+M\omega_0^2)\right]}\right|^{-\frac{1}{2}}T.
   \end{equation}
The negative eigenvalue guarantees a finite lifetime or equivalently a non-zero escape rate and characterizes the possibility for the particle to tunnel.
The function $K_j(T)$ can be evaluated from the asymptotic shape of the bounce orbit.
If we suppose $\bar{z}(T\to\infty)\sim\sigma_j\gamma_j\Exp{-\omega_0T}$, where $\gamma_j=\Exp{C_j}$ is obtained from
   \begin{equation}
   C_j=\int_0^{\sigma_j}\mathrm{d}z\left[\sqrt{\frac{M\omega_0^2}{2U(z)}}-\frac{1}{z}\right],
   \end{equation}
the Gelfand-Yaglom formula gives finally rise to
   \begin{equation}
   K_j(T)=\frac{i}{2}\sqrt{\frac{M\omega_0}{\pi\hbar}}\omega_0\sigma_j\gamma_jT.
   \end{equation}

   \begin{figure}[t]
   \includegraphics[width=8.6cm]{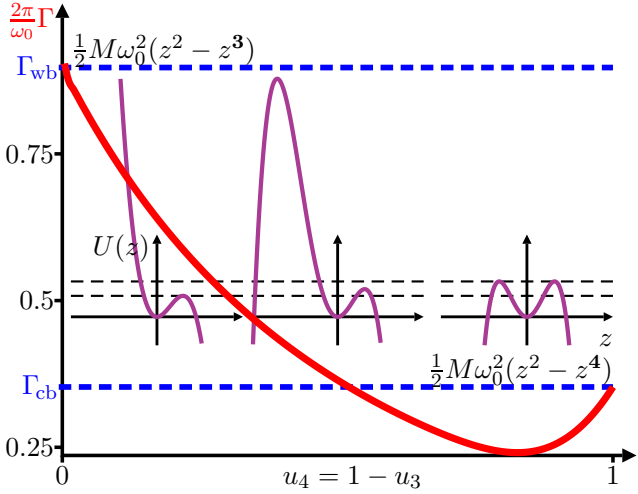}
   \caption{(Color online) Tunneling rate $\Gamma$, in red, for different barrier shapes, plotted in violet, from cubic to purely quartic.
   The rate is in units of the bottom well frequency $\omega_0/2\pi$ and the mass parameter is fixed to $M\omega_0/\hbar=7$.
   The tunneling rates Eq.~\eqref{Gwb} for $u_4=0$ and Eq.~\eqref{Gcb} for $u_3=0$ correspond to the dashed blue lines.}
   \label{figrates}
   \end{figure}

During the time $T$, several bounces occur, and the total escape rate is obtained after summing over all the possible configurations.
The multibounce configurations consist of $n_R$ right bounces and $n_L$ left bounces with centers spread between $-T/2$ and $T/2$.
We use the dilute instanton gas approximation, where the bounces are considered to be independent, which is valid when the time between two successive attempts is larger than the tunneling time.
The action of a multibounce is $n_LS_0^L+n_RS_0^R$ and the contribution from the quantum fluctuations is $K_L^{n_L}K_R^{n_R}/n_L!n_R!$.
The sum over the bounce number results in
   \begin{multline}
   G(T)=\sqrt{\frac{M\omega_0}{\pi\hbar}}\exp\!\left[-\frac{1}{2}\omega_0T\right.\\\left. + \Exp{-S_0^L/\hbar}K_L(T) + \Exp{-S_0^R/\hbar}K_R(T)\right].
   \end{multline}
The total decay rate $\Gamma$ is
   \begin{equation}
   \Gamma=\Gamma_L+\Gamma_R,
   \label{totalrate}
   \end{equation}
where
   \begin{equation}
   \Gamma_j=2\Exp{-S_{0,j}/\hbar}\mathrm{Im}K_j(T)/T.
   \end{equation}
As a conclusion, in the limit of a dilute gas of instantons, the total escape rate is
simply the sum of the tunneling rates for each barrier.
This simple result is a direct consequence of the dilute instanton gas approximation since interference effects between successive tunneling events would change the tunneling rate in one barrier and the resulting rate for two escape paths.

As a result, the tunneling rate of a particle of mass $M$ from an unstable state through
the barrier potential $U(z)$ with a double escape path reads
$\Gamma=\sum_{j=L,R}\Gamma_j$ where
   \begin{align}
   \Gamma_j&=A_j\,\Exp{-B_j/\hbar},\\
   A_j&=\omega_0\sqrt{\frac{M\omega_0}{\pi\hbar}}\,\sigma_j\,\Exp{C_j},\\
   B_j&=2\int_{0}^{\sigma_j}\textrm{d}z \sqrt{2MU(z)}.
   \end{align}

Applied to the quartic potential of Eq.~\eqref{genquarticpot}, this general result gives rise to the total tunneling rates
   \begin{align}
   \Gamma_{L,R}=&
   4\omega_0\sqrt{\frac{M\omega_0}{\pi\hbar}}\frac{\mp\sigma_{L,R}}{2-u_3\sigma_{L,R}}\nonumber\\
   \times&\exp\!\left\{-\frac{2M\omega_0}{u_4\hbar}\left[\frac{1}{3}+\frac{u_3^2}{8u_4}\right.\right.\nonumber\\
   &\left.\left.\pm u_3\frac{u_3^2+4u_4}{16u_4^{3/2}}\arccos\!\left(\frac{\mp u_3}{\sqrt{u_3^2+4u_4}}\right)\right]\right\}.
   \label{GLRgenpot}
   \end{align}
Eqs.~\eqref{totalrate} and~\eqref{GLRgenpot} constitute the central result of our work.
The MQT rate is plotted in Fig.~\ref{figrates} for a potential varying from a cubic to a
purely quartic profile.
Starting from a purely cubic potential, the tunneling rate decreases as the barrier height increases.
When the tunneling rates of both barriers are non-vanishing, the total tunneling rate is enhanced and reaches another maximum for the purely quartic potential.
This value of the tunneling rate is also obtained for a quartic potential when the cubic and quartic contributions are similar ($u_3\simeq u_4\simeq0.5$).

   \begin{figure}[t]
   \includegraphics[width=8.6cm]{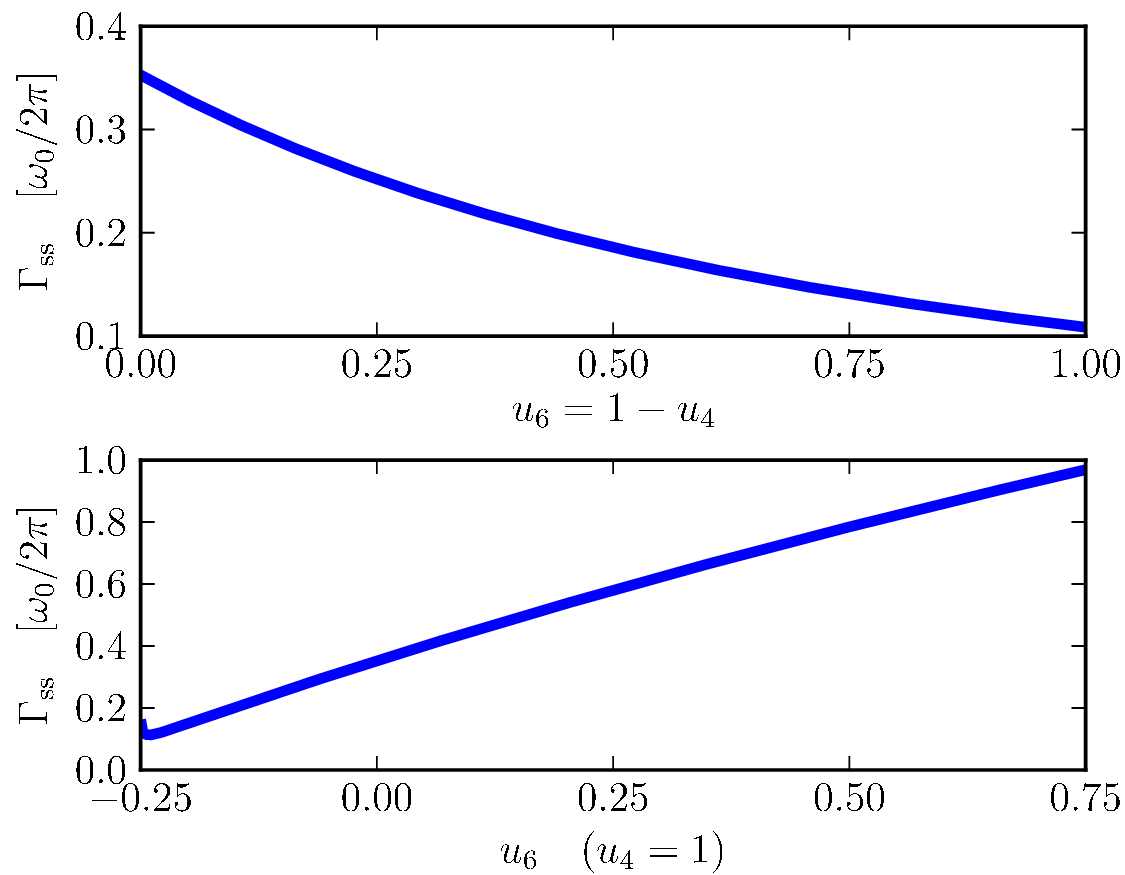}
   \caption{(Color online) Effect of a sextic contribution on the MQT rate of a quartic potential.
   The rate is in units of the bottom well frequency $\omega_0/2\pi$ and the mass parameter is fixed to $M\omega_0/\hbar=7$.}
   \label{figsextic}
   \end{figure}

When the potential is symmetric and has a sextic contribution,
$U(z)=\frac{1}{2}M\omega_0^2z^2(1-u_4z^2-u_6z^4)$, the fictitious particle can use two
paths to escape at the exit points $\pm\sigma_s$ with the same MQT rate
$\Gamma_\mathrm{ss}/2$. The exit points are defined with
$\sigma_s=\sqrt{(\sqrt{u_4^2+4u_6}-u_4)/2u_6}$ and the sextic coefficient satisfies
$u_6>-u_4^2/4$ ($u_4>0$).
At $u_6=-u_4^2/4$, the three minima are at the same height and a macroscopic quantum coherence develops between them.
The possibility to tunnel between them lifts the degeneracy and the spectrum becomes composed of three hybridized levels separated by an energy given by the tunneling strength.
Our general results can be applied to this symmetric sextic potential as well and leads to
   \begin{align}
   \Gamma_\mathrm{ss}=&
   4\omega_0\sqrt{\frac{M\omega_0}{\pi\hbar}}\frac{\sigma_s}{\sqrt{2-u_4\sigma_s^2}}\nonumber\\
   \times&\exp\!\left\{\frac{M\omega_0}{8u_6\hbar}\left[2u_4
   -\frac{u_4^2+4u_6}{\sqrt{u_6}}\arctan\!\left(\frac{2\sqrt{u_6}}{u_4}\right)\right]\right\}.
   \label{Gsextic}
   \end{align}
In Appendix~\ref{apptopo} we show that the camel-back potential is in fact better fitted with a sextic contribution.
The resulting MQT rate is presented in Fig.~\ref{figsextic}.
The sextic contribution mainly sharpens the camel-back potential and hence increases the tunneling rate.

The tunneling rate of the ground state can be used to calculate the tunneling rate for
the excited states~\cite{footnote}.

\section{Washboard and camel-back potentials}

   \begin{figure}[t]
   \centering
   \includegraphics[width=8.6cm]{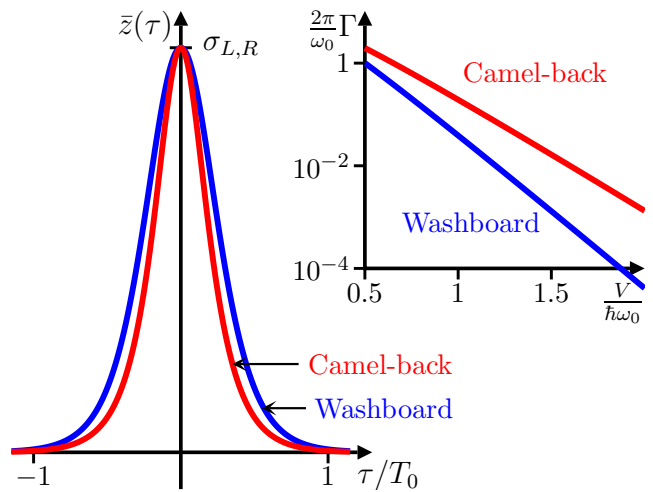}
   \caption{(Color online) Bounce orbits, Eqs.~\eqref{Zwb} and~\eqref{Zcb}, and MQT rates, Eqs.~\eqref{Gwb} and~\eqref{Gcb}, of the washboard potential and the camel-back potential.
   The tunneling time is of the order of the period $T_0$ in the bottom well and, for the same bottom well frequency and barrier height, the escape rate is larger for the camel-back potential than for the washboard potential.}
   \label{figorbit}
   \end{figure}

Applied to the washboard potential ($u_4=0$), the standard MQT tunneling rate is recovered~\cite{Kagan}
   \begin{equation}
   \Gamma_\mathrm{wb}=12\sqrt{3}\,\frac{\omega_0}{2\pi}\sqrt{\frac{2\pi V}{\hbar\omega_0}}\,\exp\!\left[-\frac{36}{5}\frac{V}{\hbar\omega_0}\right],
   \label{Gwb}
   \end{equation}
where $V=2M\omega_0^2/27c_3^2$.
From Eq.~\eqref{eqbounceorbit}, the bounce orbit is found to be
   \begin{equation}
   \bar{z}_\mathrm{wb}(\tau)=\frac{\sigma_{L,R}}{\cosh^2(\omega_0\tau/2)}.
   \label{Zwb}
   \end{equation}

In the case of the camel-back potential ($u_3=0$), we get the tunneling rate
   \begin{equation}
   \Gamma_\mathrm{cb}=16\,\frac{\omega_0}{2\pi}\sqrt{\frac{2\pi V}{\hbar\omega_0}}\,\exp\!\left[-\frac{16}{3}\frac{V}{\hbar\omega_0}\right],
   \label{Gcb}
   \end{equation}
where $V=M\omega_0^2/8c_4$.
Eq.~\eqref{Gcb} can also be obtained from the symmetric sextic MQT rate Eq.~\eqref{Gsextic} with $u_6=0$.
The bounce orbit of the camel-back potential is equal to
   \begin{equation}
   \bar{z}_\mathrm{cb}(\tau)=\frac{\sigma_{L,R}}{\cosh(\omega_0\tau)}.
   \label{Zcb}
   \end{equation}

As seen in Sec.~\ref{phasequbit}, these two regimes can be observed in a phase qubit close to the critical line and around the working point biased with a weak current and half a flux quantum, respectively.
The bounce orbits as well as the MQT rates are plotted in Fig.~\ref{figorbit}.
The tunneling time turns out to be of the order of the period of oscillation $T_0=2\pi\omega_0^{-1}$.
The approximation of a dilute gas of instantons is valid when the time between two tunneling events $\Gamma^{-1}$ is larger than the tunneling time $T_0$, i.e. $\Gamma<\omega_0/2\pi$.
This corresponds to a barrier height larger than the ground state energy $\hbar\omega_0/2$.
The escape rate from the camel-back potential of parameters $(\omega_0,\,V)$ corresponds to the escape rate from a cubic potential of parameters $(\frac{2}{\sqrt{5}}\omega_0\simeq0.89\,\omega_0,\,\frac{8\sqrt{5}}{27}V\simeq0.66\,V)$.
For two potentials of the same frequency and same barrier height, obtained when $c_3=\frac{4}{3\sqrt{3}}\sqrt{c_4}$, the ratio of the escape rates is $\Gamma_\mathrm{cb}/\Gamma_\mathrm{wb}=\frac{4}{3\sqrt{3}}\Exp{\frac{28}{15}\frac{V}{\hbar\omega_0}}\simeq0.77\,\Exp{1.9\frac{V}{\hbar\omega_0}}$.
The tunneling rate out of the camel-back potential is then larger than the escape rate from the washboard potential when $V/\hbar\omega_0>\frac{15}{28}\log\frac{3^{3/2}}{4}\simeq0.14$, see Fig.~\ref{figorbit}.

Finally, for a purely sextic potential $U(z)=\frac{1}{2}M\omega_0^2z^2(1-u_6z^4)$, the total rate reads
   \begin{equation}
   \Gamma_\mathrm{ps}=\frac{12}{\sqrt[4]{3}}\frac{\omega_0}{2\pi}\sqrt{\frac{2\pi V}{\hbar\omega_0}}\,\exp\!\left[-\frac{3\sqrt{3}\pi}{4}\frac{V}{\hbar\omega_0}\right],
   \label{Gps}
   \end{equation}
where the barrier height is equal to $V=M\omega_0^2/3\sqrt{3u_6}$ for $u_6>0$.

\section{Conclusion}

Using the instanton technique we calculate the escape rate from the metastable state of a quartic and a symmetric sextic potential, in which the fictitious particle can tunnel through two different barrier potentials.
This work is of interest for Josephson junction based nanocircuits, such as in the field of circuit quantum electrodynamics, where the potential shapes are designed for an efficient quantum information processing.
In particular, the results can be directly used in the case of phase qubits, where we give the tunneling rate as a function of the biases in two different regimes, namely close to the critical line and also close to the working point with a weak current bias and half a flux quantum magnetic bias.

\begin{acknowledgements}
We thank O.~Buisson, E.~Hoskinson, and F.~Lecocq for the opportunity to interact with the experiment of Ref.~\onlinecite{camelback}.
N.~D. thanks B.~Dou\c cot for his useful comments on the work.
\end{acknowledgements}

\appendix

\section{Topography of the phase qubit}
\label{apptopo}

In this Appendix we present the shape of the one-dimensional potential, \textit{i.e.} $\omega_0$, $u_3$, and $u_4$, of a phase qubit in two different regimes, namely the washboard potential and the camel-back potential.
The bottom well frequency and the barrier height can be directly used to obtain the MQT rate as a function of the experimental parameters.

\subsection{Washboard potential}

Close to the critical line the one-dimensional potential has a washboard shape.
We Taylor expand the potential in $x$, $y$ and $s$ at third order around $x_c$, $y_c$, and $s_c$ in the direction of vanishing curvature, given by the angle
$\tan\theta_c=-\partial^2_{xx}u(x_c,y_c)/\partial^2_{xy}u(x_c,y_c)$.
The potential can be expressed in terms of the barrier height $V$ and the bottom well frequency $\omega_0$ as follows
   \begin{align}
   U(z)=&\frac{1}{2}M\omega_0^2z^2\left(1-\sqrt{\frac{2M\omega_0}{27V}}z\right),\\
   V=&\frac{4}{3}\left(\frac{2s_c}{c_3}\right)^{1/2}(\cos\theta_c+\eta\sin\theta_c)^{3/2}(1-s)^{3/2}E_J,\\
   \omega_0=&\left(\frac{c_3}{2s_c}\right)^{1/4}(\cos\theta_c+\eta\sin\theta_c)^{1/4}(1-s)^{1/4}\omega_p,
   \end{align}
where $c_3=\cos\theta_c(1+2\sin^2\theta_c)(\sin x_c\cos y_c-\alpha\cos x_c\sin y_c)+\sin\theta_c(1+2\cos^2\theta_c)(\cos x_c\sin y_c-\alpha\sin x_c\cos y_c)$.
These expressions generalize the known results for a symmetric SQUID~\cite{lefevreseguin} ($\eta=0$)
and can be directly used in Eq.~\eqref{Gwb} to obtain the MQT rate.

   \begin{figure}[t]
   \centering
   \includegraphics[width=8.6cm]{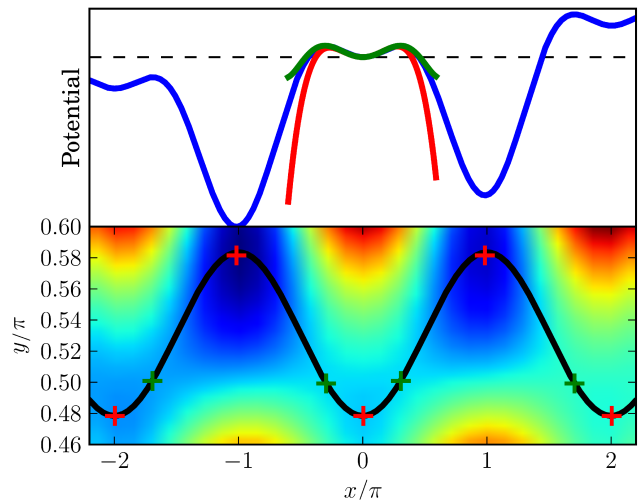}
   \caption{(Color online) Potential (top panel) along the path of minimum curvature $y=y_B-\cos(x)/2b$ (bottom panel),
   for a dc SQUID with parameters $\alpha=0.0072$, $\eta=0.69$, $b=3.03$, and the biases $s=-0.0062$ and $y_B=0.531\pi$.
   Top panel: quartic and symmetric sextic expansions, in red and green respectively.
   Bottom panel: potential $u(x,y)$ and position obtained from numerics of the minima and saddle points, red and green crosses, respectively.}
   \label{figpotpath}
   \end{figure}

\subsection{Camel-back potential}

We focus on the domain of bias fields around the critical point $\{s_c=-\alpha,{y_B}_c=\pi/2+1/2b,x_c=0,y_c=\pi/2\}$, the others can be obtained by symmetry.
The path of minimum curvature follows the trajectory $y=y_B-\cos(x)/2b$.
The Tailor expansion of the one-dimensional potential up to fourth order in $z$ reads $U(z)=\tfrac{1}{2}M\omega_0^2z^2(1-u_3z-u_4z^2)$ with
   \begin{equation}
   \omega_0=\omega_p\sqrt{c_2},\quad
   u_3=-\frac{c_3}{3c_2},\quad
   u_4=-\frac{c_4}{12c_2},
   \end{equation}
where
   \begin{align}
   c_2&=-\frac{1+\eta s-\cos\beta}{2b}-\sin\beta,\\
   c_3&=\alpha\cos\beta+\frac{3\alpha}{2b}\sin\beta,\\
   c_4&=\frac{4+\eta s-7\cos\beta}{2b}+\left(1-\frac{3}{4b^2}\right)\sin\beta,
   \end{align}
and $\beta=y_B-{y_B}_c$.
The path of minimum curvature and the corresponding potential with a quartic and a symmetric sextic expansions are plotted in Fig.~\ref{figpotpath}.
The sextic contribution is necessary to fit accurately the potential between the two exit points.


\begin{thebibliography}{9}
\bibitem{footnote}
The escape rate $\Gamma_n=\sum_{j=L,R}\Gamma_{j,n}$ of the $n^\mathrm{th}$ excited state, with energy~$E_n$, is
$
\Gamma_{j,n}=\Gamma_{j,0}\,\frac{T_0}{T_n}\frac{(T_0A_j)^{\chi_n}}{(2\pi)^{\chi_n}\chi_n!},
$
where we note $\chi_n=E_n/\hbar\omega_0-1/2$ and $T_n\simeq2\pi/\omega_0$ is the period of oscillation in the well of the $n^\mathrm{th}$ state~\cite{Weiss83}.
\end{thebibliography}
\end{document}